%% file: main.tex
\documentclass[sigconf]{acmart}
\setcopyright{rightsretained}

\usepackage{booktabs}
\usepackage{color}
\usepackage{graphicx}
\usepackage{amsthm}
\usepackage{caption}
\usepackage{subcaption}
\usepackage{balance}
\usepackage{hyperref}

\usepackage{setspace}

\theoremstyle{definition}
\newtheorem{example}{Example}

\begin{document}

\title{\textit{Can Women Break the Glass Ceiling?}: An Analysis of \#MeToo Hashtagged Posts on Twitter}

\author{Naeemul Hassan}
\affiliation{%
  \institution{University of Mississippi}
}
\email{nhassan@olemiss.edu}

\author{Manash Kumar Mandal}
\affiliation{%
  \institution{Khulna University of Engineering and Technology}
}
\email{manashmndl@gmail.com}

\author{Mansurul Bhuiyan}
\affiliation{%
  \institution{IBM Research, Almaden}
}
\email{mansurul.bhuiyan@ibm.com}

\author{Aparna Moitra}
\affiliation{%
  \institution{University of Delhi}
}
\email{aparna.moitra@gmail.com}

\author{Syed Ishtiaque Ahmed}
\affiliation{%
  \institution{University of Toronto}
}
\email{ishtiaque@cs.toronto.edu}

\renewcommand{\shorttitle}{\textit{Can Women Break the Glass Ceiling?}: \#MeToo}

\renewcommand{\shortauthors}{Anonymous}

\begin{abstract}
    In October 2017, there happened the uprising of an unprecedented online movement on social media by women across the world who started publicly sharing their untold stories of being sexually harassed along with the hashtag \#MeToo (or some variants of it). Those stories did not only strike the silence that had long hid the perpetrators, but also allowed women to discharge some of their bottled-up grievances, and revealed many important information surrounding sexual harassment. In this paper, we present our analysis of about one million such tweets collected between October 15 and October 31, 2017 that reveals some interesting patterns and attributes of the people, place, emotions, actions, and reactions related to the tweeted stories. Based on our analysis, we also advance the discussion on the potential role of online social media in breaking the silence of women by factoring in the strengths and limitations of these platforms.
\end{abstract}

\maketitle

\input{sec-introduction}

\input{sec-litreview}
\input{sec-data}
\input{sec-analysis}
\input{sec-limitations}
\input{sec-discussion}

\small
\Urlmuskip=0mu plus 1mu
\bibliographystyle{ACM-Reference-Format}
\bibliography{main}

\end{document}

%% file: sec-introduction.tex
\section{Introduction}
\label{sec-introduction}

\begin{quote}
    \textit{``I was 16. My middle aged male boss harassed me. I never talk about it. He wasn't the last."}
\end{quote} 

This and many such tweets started to explode the Twitter news-feed soon after actress Alyssa Milano encouraged spreading the \#MeToo phrase as part of an awareness campaign in order to reveal the ubiquity of sexual harassment, tweeting: \textit{``If all the women who have been sexually harassed or assaulted wrote `Me too' as a status, we might give people a sense of the magnitude of the problem"} on October 15, 2017 {\let\thefootnote\relax\footnote{{*All the URLs in footnote taking more than a line have been shortened to save space}}} ~\footnote{http://bit.ly/boston-globe-metoo-origin}. However, the origin of \#MeToo dates back to 2006 when social activist Tarana Burke launched a grass-root level campaign for ``empowering through empathy" for the women of color within their community ~\footnote{http://bit.ly/independent-metoo-why-when}. Milano's call for sharing harassment experiences with \#MeToo hashtag that followed her own allegation against Harvey Weinstein, an American film producer, for sexually abusing her ~\footnote{http://motto.time.com/4987331/alyssa-milano-me-too-sexual-assault/} took the original movement to a whole new level and millions of women around the world started participating. Before this, a few other hashtags were also used for similar purposes ~\footnote{http://bit.ly/metoo-other-tags}, including \#MyHarveyWeinstein, \#YouOkSis, \#WhatWereYouWearing, and \#SurvivorPrivilege. However none of them could create such a massive movement on social media. 

After Milano's tweet, within 24 hours, there were more than 4.5 million posts on different social media with hashtag \#MeToo (or some variants of it) ~\footnote{http://bit.ly/usnews-metoo}. She got more than 70 thousand replies to her tweet on Twitter ~\footnote{http://fortune.com/2017/10/17/me-too-hashtag-sexual-harassment-at-work-stats/} in one day. The hashtag quickly became viral and in the next few days, there were millions of posts on Twitter, Facebook, Instagram, and other social media where victims of sexual harassment were sharing their experiences, revealing the name of their harassers, accusing institutions for not being strict about harassment, and reflecting on different laws and policies. While majority of the victims were women, there were also victims of other genders. Moreover, many other users joined this movement by posting with this hashtag to show their support to the cause. Women from all over the world, women of different age groups, of different colors, and of different professions shared their experiences of being harassed at their home, at their workplace, at public places, and over social media. These hashtagged posts were also shared, re-posted, and re-tweeted. The silence around sexual harassment was broken, and the tabooed topics of ``sex" and ``harassment" were widely discussed across the world.

Millions of tweets that women shared in public revealed many unknown stories of famous personalities, including politicians, artists, film stars, professors, and businessmen. Such stories challenged many long-held beliefs and perceptions regarding various people, places, and professions. For example, many Hollywood actresses, producers, writers, and directors revealed a dirty facet of this film industry where many women are regularly abused ~\footnote{http://bit.ly/rollingstone-alyssa-milano}. Their voice led them toward building a new initiative called \textit{Time's Up} that aims to combat sexual harassment in Hollywood and in blue-collar workplaces nationwide ~\cite{nytimesmetoo}. Female veterans also joined \#MeToo movement and revealed their struggles against sexual harassment in military service ~\footnote{https://firenewsfeed.com/news/1024194}. Students and professors also joined \#MeToo movement, and revealed how sexual harassment take place in an academic setting ~\footnote{http://bit.ly/foxnews-universities-metoo}. Surprising and touching stories of sexual harassment in K-12 schools also started to come to the light along with \#MeToo (later this became a separate movement called  \#MeTooK12 ~\footnote{http://bit.ly/wpost-metoo12k}). The prevalence and severity of sexual harassment that came to the light through those stories shook the world with disbelief, disgust, and horror. 

This world-wide online revolution of \#MeToo did not only initiate a movement against violence, assault, and harassment, but also raised some very important scholarly questions relevant to the Hypertext community and Computer Science discipline in general. In this paper, we ask, 
\begin{enumerate}
    \item Who are sharing harassment stories on Twitter?
    \item What kind of new knowledge could we gather around sexual harassment through these tweets?
\end{enumerate}
By making an attempt to answer these questions, in this paper, we also explore the role of online social media in empowering women by providing them with a platform for raising their voice.

%% file: sec-litreview.tex
\section{Related Work}
\label{Related Work}

Sexual harassment is bullying or coercion of a sexual nature, or the unwelcome or inappropriate promise of rewards in exchange for sexual favors~\cite{paludi1991academic}. This is a serious offence that causes several negative impacts on the victims' physical, psychological, and social health~\cite{fitzgerald1997antecedents, glomb1997ambient, loy1984extent}. Unfortunately, the prevalence of sexual harassment is very high across the globe~\cite{gelfand1995structure, paludi1991academic}, and women are the principle victims of sexual harassment. According to UN Women and World health Organization (WHO), 35 percent (almost one in every three) of women worldwide have experienced either physical and/or sexual intimate partner violence or sexual violence by a non-partner at some point in their lives ~\footnote{http://bit.ly/unwomen-violence-facts}. A 2017 survey says, one-fifth of American adults have experienced sexual harassment at their workplace ~\footnote{http://bit.ly/cnbc-harassment-at-work}. The situation has been found to be similarly alarming in the UK ~\footnote{http://bit.ly/telegraph-harassment-at-work} and Australia ~\footnote{http://bit.ly/gov-australia-sexual-harassment}. The national crime statistics agency of France published a report in December, 2017 that said more than 220 thousand women were harassed in the public transport facilities of the country ~\footnote{http://bit.ly/reuters-france-public-transport-harassment}. Such statistics are often hard to get in many places in Africa and Asia, but several studies and anecdotal evidences demonstrate the wide prevalence and severity of sexual harassment in \footnote{http://bit.ly/timesindia-harassment-india}~\footnote{ https://africacheck.org/factsheets/guide-rape-statistics-in-south-africa/}~\footnote{https://unstats.un.org/unsd/gender/chapter6/chapter6.html} those continents, too. 
All these alarming reports show us how women are being sexually abused at their home, at their workplace, and in public places by their husbands, boyfriends, colleagues, professors, and even by strangers. 

However, most incidents of sexual harassment remain unreported~\cite{arvey1995using}. Many women do not talk about the sexual harassment that they experience~\cite{ahmed2014protibadi}. In the western world, studies show that such silence is sometimes caused by the fear of retaliation~\cite{bergman2002reasonableness, fitzgerald1995didn} - the fear that the allegation could be trivialized and ignored, or could even backfire. In some other cases, women do not report because they cannot imagine other people helping them, which is known as "bystander effect"~\cite{banyard2011will}. Moreover, many workplaces harbor a ``masculine culture" where harassment is seen as an act of power, and the female workers often prefer to keep silent about the harassment they experience in order to fit in and be "one of the guys"~\cite{fitzgerald1993sexual}. Speaking out is even harder for women in many places in the global south. In most places in the Indian subcontinent, for example, ``sex" is a tabooed topic, and talking about sexual harassment is seen as an act of immodesty~\cite{nair2000question}. This challenge is compounded by the strong patriarchal culture~\cite{houston1991speaking} in India and the middle-east~\cite{shalhoub2003reexamining}, where women often depend on men for most things needed for their living including foods, shelter, education, transportation, and health~\cite{mitter1992dependency}. Furthermore, feminist movements there are often associated with western culture~\cite{moghadam2005globalizing, mohanty1988under} that many nations have a strong detest for due~\cite{brenner2003transnational} to their long and painful colonial past. Postcolonial scholar, Gayatri Spivak, famously asked, \textit{``Can the subaltern speak?"} to emphasize on the constraints that keep a woman silent in such contexts~\cite{morris2010can}. For these and may other reasons, most women remain silent about the sexual harassment that they experience.


Feminist activists, researchers, artists, and scholars have long been trying to break this silence of women about sexual harassment. There have been many demonstrations on the street and on the news, radio, and television media by the feminist activists that depict the severity of sexual harassment, and the gender-politics related to it~\cite{hester1996women}. With the advancement of information and communication technologies, people have now many effective means of communication including mobile phone and internet. Researchers have tried to leverage them to design different communication platforms for women to report abuses, get safety information, combat harassment, and share their opinions with others. For example, \textit{Hollaback!} is a smartphone application that allows the users to take pictures of the harassers and post the pictures on social media to shame them. Women can also report harassment through \textit{Harassmap}, an application that shows an interactive map of sexual harassment to the users~\cite{young2014harassmap}. \textit{Safetipin} is another mobile phone application with which users can mark the unsafe places, and it also informs a user if they enter into an unsafe place~\cite{viswanath2015safetipin}. On the other hand, \textit{Circle of 6} helps users find other trusted women to accompany them while traveling~\footnote{\url{ https://www.circleof6app.com}}. In Bangladesh, \textit{Protibadi}, a mobile application was deployed that allowed women to anonymously report harassment, provided an interactive harassment map, and allowed the users to participate in discussion around harassment anonymously~\cite{ahmed2014protibadi}. All these applications have been proven to be effective to some extent, and could help some women break the silence and raise their voice against sexual harassment. However, as it has been seen, many women do not use these platforms to talk about sexual harassment because of many larger social, cultural, and political challenges that just cannot be fully mitigated only by easy, accessible, ubiquitous, and intelligent applications and communication~\cite{ahmed2016computing} tools. Moreover, without the spontaneous participation of all women, getting relevant information to improve the design of an effective technology remains challenging, too.

The \#MeToo movement on Twitter and other social media is hence considered revolutionary. Such voluntary and spontaneous participation of millions of women across the globe in breaking down a silence that had long been suppressing them was unprecedented in the history. Through the tweets of these women, we could know many information of sexual harassment that were never disclosed before. We could also learn the context, emotion, and repercussions around sexual harassment which would be hard to get otherwise. We argue that the tweets shared by the victims over Twitter have not only helped them release their bottled up emotions, but also have provided us with important data that we can use for better understanding the nature of sexual harassment. Such information can be used for improving existing law, policy, education, and technology to reduce the occurrences of sexual harassment and to extend a better support to the victims. 

Studying \#MeToo movement is also important to understand the role of social media in activism, movements, and politics. Several studies~\cite{shirky2011political} have been conducted to analyze how social media played an important role in many recent political movements around the world including Arab Spring, Shahbag movement, Brazilian protest, and Brexit. Researchers have found that social media helped the organization of those movements~\cite{juris2005new} by quickly disseminating information and strengthened solidarity by propagating emotion~\cite{rahimi2011agonistic}. At the same time, such movements brought to the fore the commonly shared values~\cite{carroll2006democratic} of the protesting communities and many social leaders were born~\cite{gerbaudo2012tweets}. A big and important part of those movements took place in the streets, and social media played a crucial role in making the protest stronger. Different other movements on social media are different in nature than those political movements. For example, activism around climate change on social media is often targeted to aware and educate people of the potential harm of a policy, law, or technology~\cite{segerberg2011social}. \#blacklivesmatter, a hashtag movement on social media to protest racial discrimination and violence against the African American communities have been acting as a powerful tool to put a surveillance upon the police actions and government decisions in the US~\cite{garza2014herstory}. Such movements that are targeted to a long-term social change put surveillance over the society around an issue, educate people of their concerns, keep people aware of new incidents relevant to the movement, and develop a community that further the movement both in the virtual and `real' worlds. We argue that \#MeToo falls under this category of long-term social movements, and we need to study its role in educating, spreading information, and developing community to better understand its potential impact on the society.

%% file: sec-data.tex
\section{Data Collection}
\label{sec-data}

We collected about a million tweets containing the \emph{\#MeToo} hashtag from all over the world within the October $15-31$, $2017$ period using the Twitter API. We removed the duplicate tweets from this collection by matching the permanent URLs of the tweets. After that, we removed the non-English language tweets, and re-tweets of original tweets. These cleaning steps reduced the collection size to $441,925$. These tweets originated from $323,813$ unique users. A portion ($9,580$) of this collection has geolocation. Table \ref{tab-city-gender} shows distribution of the geolocated data among the cities. We take a $25\%$ random sample from the whole dataset and make it publicly ~\footnote{http://bit.ly/metoo\_data} available for the academicians. Note, rather than sharing the original tweets, we release the unigram frequencies for each tweet. 

\begin{table}[t!]
\centering
\vspace{-0.12in}
\caption{Distribution of geolocated tweets among the cities}
\label{tab-city-gender}
\vspace{-0.12in}
\begin{tabular}{l|rrrr}
\toprule
city         & \#Tweets & Male & Female & Female(\%) \\
\midrule
Dallas       & 1249     & 144  & 471    & 76.59      \\
Dhaka        & 82       & 32   & 22     & 40.74      \\
Indianapolis & 1203     & 180  & 452    & 71.52      \\
Kansan City   & 448      & 61   & 208    & 77.32      \\
Karachi      & 250      & 44   & 55     & 55.56      \\
Mumbai       & 2778     & 724  & 674    & 48.21      \\
New York      & 1878     & 267  & 754    & 73.85      \\
North Dakota  & 110      & 18   & 34     & 65.38      \\
Portland     & 849      & 146  & 316    & 68.4       \\
Saint Louis   & 619      & 92   & 224    & 70.89      \\
Tehran       & 114      & 13   & 32     & 71.11      \\
\bottomrule
\end{tabular}
\vspace{-0.12in}
\end{table}

%% file: sec-analysis.tex
\section{Analysis}
\label{sec-analysis}

\subsection{Age and Gender Distribution}
\label{sec-age-gender}


Using the Face$^{++}$ \footnote{https://www.faceplusplus.com/} API, we determined the gender and age of the Twitter users of the geolocated dataset by using their profile pictures. Some of the profile pictures do not contain any face and some of them contain multiple faces. We could identify age and gender of $4,963$ users. Among them, $3,242$ $(65\%)$ are \emph{females} and $1,721$ $(35\%)$ are males. Table ~\ref{tab-city-gender} shows gender distribution of users for each city. This distribution shows that the percentage of \emph{Female} users is comparatively lower in Asian cities (average $49.06\%$) than in U.S. cities (average $73.03\%$). This finding may seem counter-intuitive at first, but can be explained by the overwhelming gender bias in the Internet access in those Asian countries ~\cite{internetsociety}. Several studies have reported how free access to digital technologies is challenging for women in many Asian countries~\cite{ahmed2017privacy}. This lack of access to digital technologies for women in those places is often attributed to the patriarchal politics of dis-empowering them. Moreover, a close examination of the tweets that are posted by men revealed that most of them were not sharing their harassment experiences, but they were showing their support to the movement.

\begin{figure}[t!]
\centering
\includegraphics[width=0.95\linewidth]{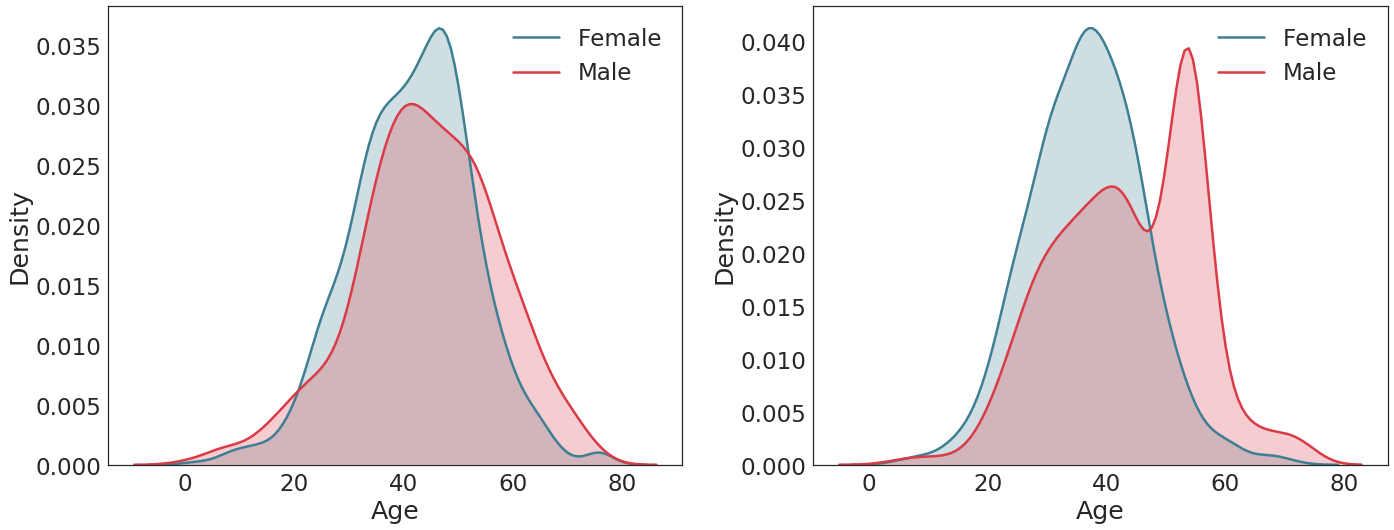}
\vspace{-0.08in}
\caption{Age distribution of female and male users in two continents (left: United States, right: Asia).}
\vspace{-0.15in}
\label{fig-age-distribution}
\end{figure}

Figure ~\ref{fig-age-distribution} shows the distribution of age of different genders in the cities in United States (left) and in Asia (right). Although the density plots for \emph{Female} in both regions are similar, a difference in \emph{Male} density plots is clear. We observe a higher participation of elderly males from Asian cities than from in U.S. cities. This finding may also be attributed to the ownership and dependency model in Asian cities. Several studies have shown that most people in Asia live a ``collective" live with their families and communities, which is different from the ``individualistic" life-style in the US and many other countries in the West~\cite{hofstede1984cultural}. Individuals often depend on the money from their parents for a longer period of time~\cite{lahiri1989dynamics} in Asia, and their parents often limit their use of Internet during this period of time. Middle-age males, who predominantly represent the chief of a household, on the other hand, own almost everything in the house, and hence their access to Internet is easier and freer than the other members of the family~\cite{katz1997intra}. 


Taken together, we see that the age and gender distribution of the collected tweets does not only represent the demographic information of the participants, but also highlights the differences in the social and cultural dynamics around Internet usage around the globe. At the same time, if we focus on the tweets from the women, we see that most of them were coming from women within the age range of 30 to 50. We found less information from women above and below this age range. While this can be explained by the demography of Internet usage, we should also keep in mind that these tweets may convey less information about the harassment happened to younger or older females - both in the US and in Asia.

\subsection{Harassment Category and Role Analysis}
This dataset gives an unprecedented opportunity to statistically analyze the categories of sexual harassment, their extents, and the roles/relation of the attackers. For each tweet, we used natural language processing to identify the presence of a sexual harassment claim and relation or role of the attacker if that is expressed. Specifically, we leveraged semantic role labeling (SRL) techniques ~\cite{carreras2005introduction} to detect claims, their subjects, verbs, and predicates. 
We used a deep highway BiLSTM with constrained decoding based SRL detection technique ~\cite{he2017deep} that achieves the best performance so far on two benchmark datasets-- $83.2\%$  F-1  on  the
CoNLL  2005 ~\cite{carreras2005introduction}  test  set  and  $83.4\%$  F-1  on
CoNLL 2012 ~\cite{pradhan-etal-conll-st-2012-ontonotes}. It provides \emph{(subject, verb, predicate)} tuple for each claim present in a text. 

\begin{example}
\emph{When I was 18 years old. I had a jerk grab my vagina at a night club in NYC.} This tweet contains multiple claims. The SRL detection system returns the tuples \emph{(I, had, a jerk grab my vagina at a night club in NYC)} and \emph{(a jerk, grab, my vagina)}. In the second claim, the role \emph{a jerk} is responsible for the action event \emph{grab}. 
\end{example}

\begin{table*}[t]
\centering
\caption{Examples of harassment claims in \emph{(subject, verb, predicate)} form and their corresponding tweets.}
\vspace{-0.12in}
\label{tab-example}
\resizebox{\textwidth}{!}{%
\begin{tabular}{llll}
\toprule
Subject & Verb & Predicate & Tweet\\
\midrule
Men                     & Catcall & me         & Men used to catcall me while I waited for mum to pick me up. I was 12 years old. \#metoo                                                     \\
Team Doctor Years.      & Molest  & she        & McKayla Maroney reveals she was molested by US gymnastic team doctor for SEVEN YEARS \#MeToo                                                 \\
Neighbor Friend Parents & Assault & i          & \#MeToo at a young age. I was assaulted by a neighbor friend of my parents. No one would believe me. Then later @18 by my Boss. I quit.      \\
College Prof. Men       & Grab    & me b4      & I've had college prof. \& men I've worked w/ grab me B4 just made it clear that wasn't happening avoided them later no repercussions \#MeToo \\
Sex Harassment Charges  & Hit    & silicon valley & \#MeToo sex harassment charges hit Silicon Valley; Robert Scoble says 'I feel ashamed'  \\
\bottomrule
\end{tabular}%
}
\end{table*}

\begin{figure}[t!]
\centering
\includegraphics[width=0.9\linewidth]{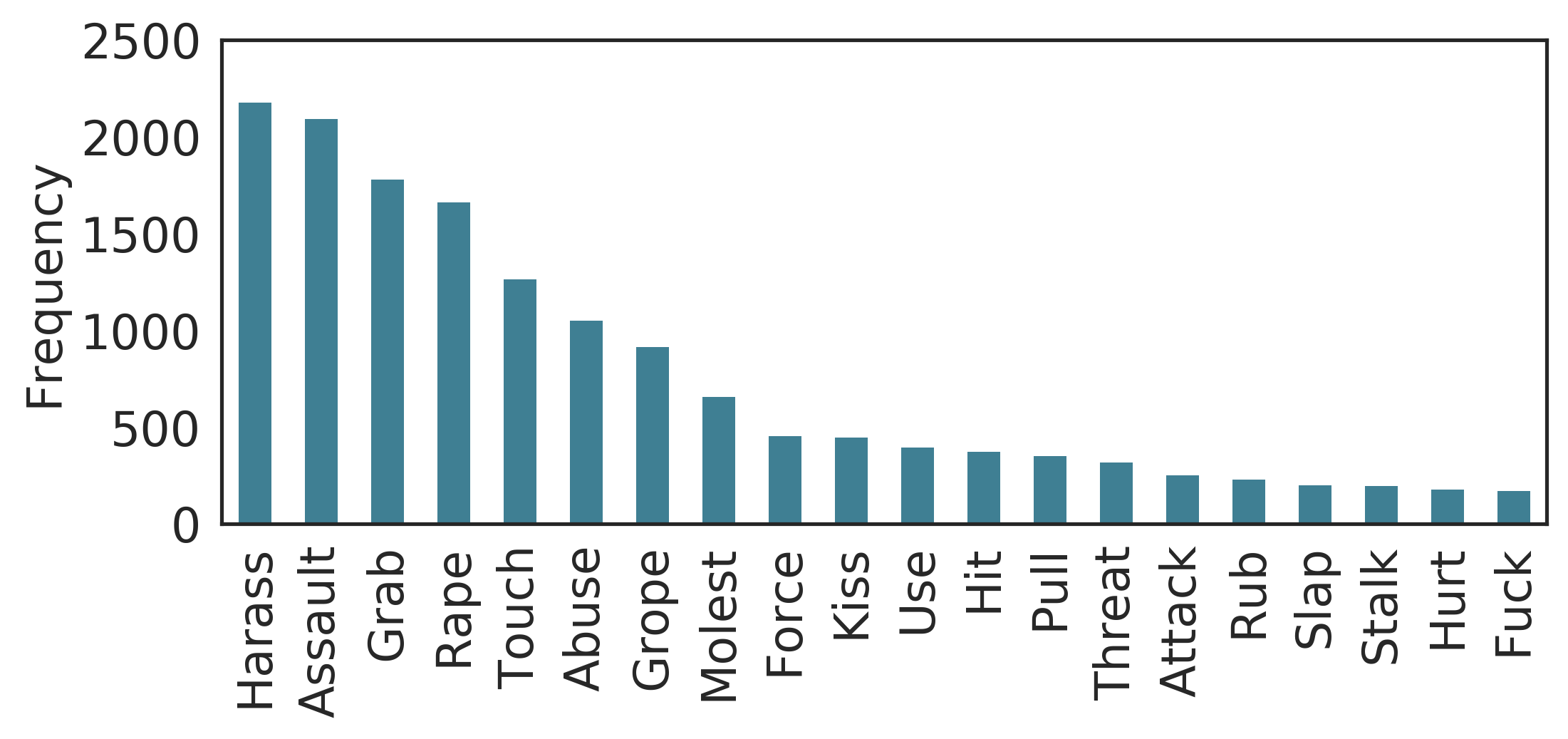}
\vspace{-0.12in}
\caption{Frequency of different harassment categories.}
\label{fig-verb-frequency}
\end{figure}

\begin{figure}
\centering
\begin{subfigure}{0.5\linewidth}
  \centering
  \includegraphics[width=\linewidth]{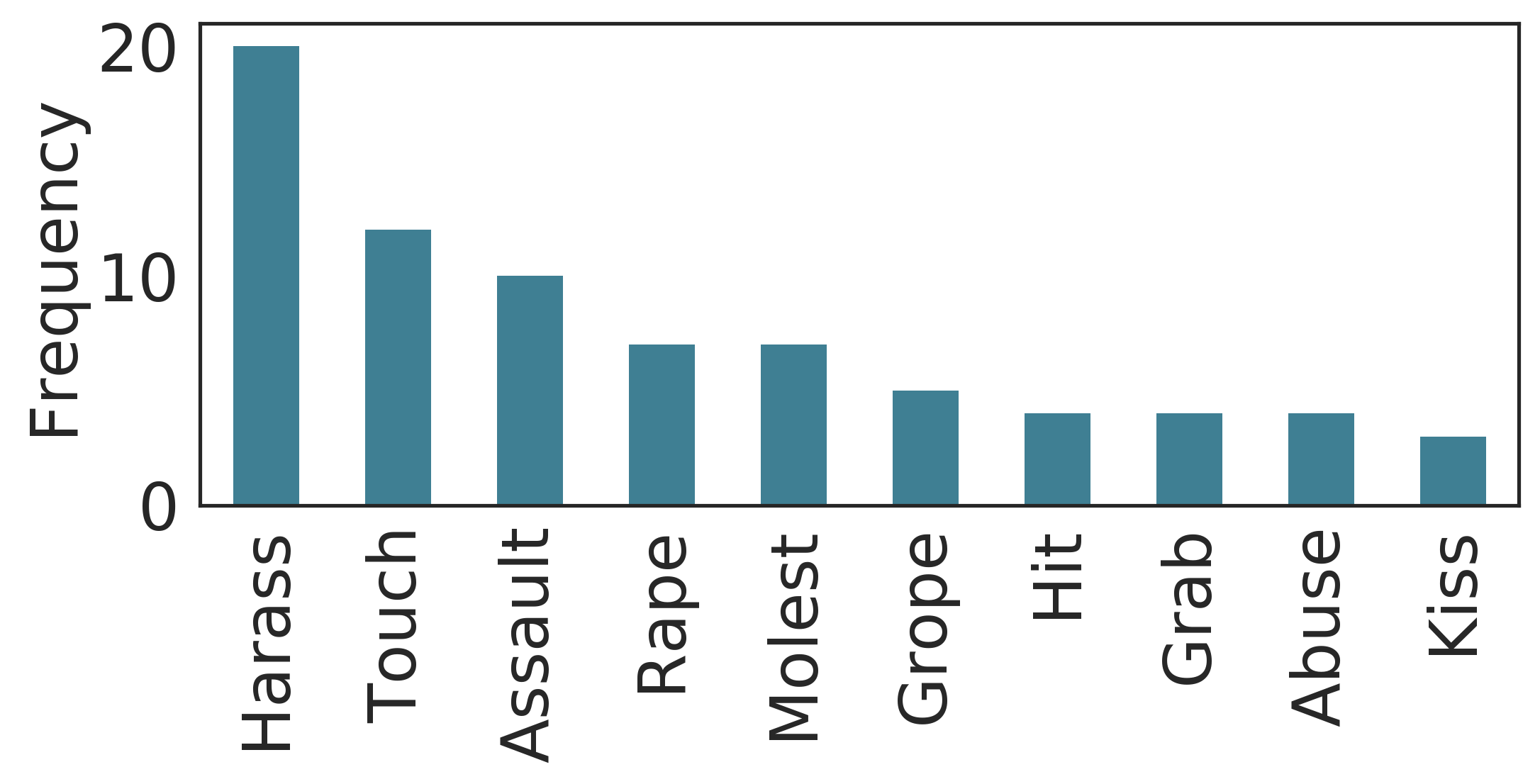}
  \caption{Asian Cities}
  \label{fig-verb-frequency-asia}
\end{subfigure}%
\begin{subfigure}{0.5\linewidth}
  \centering
  \includegraphics[width=\linewidth]{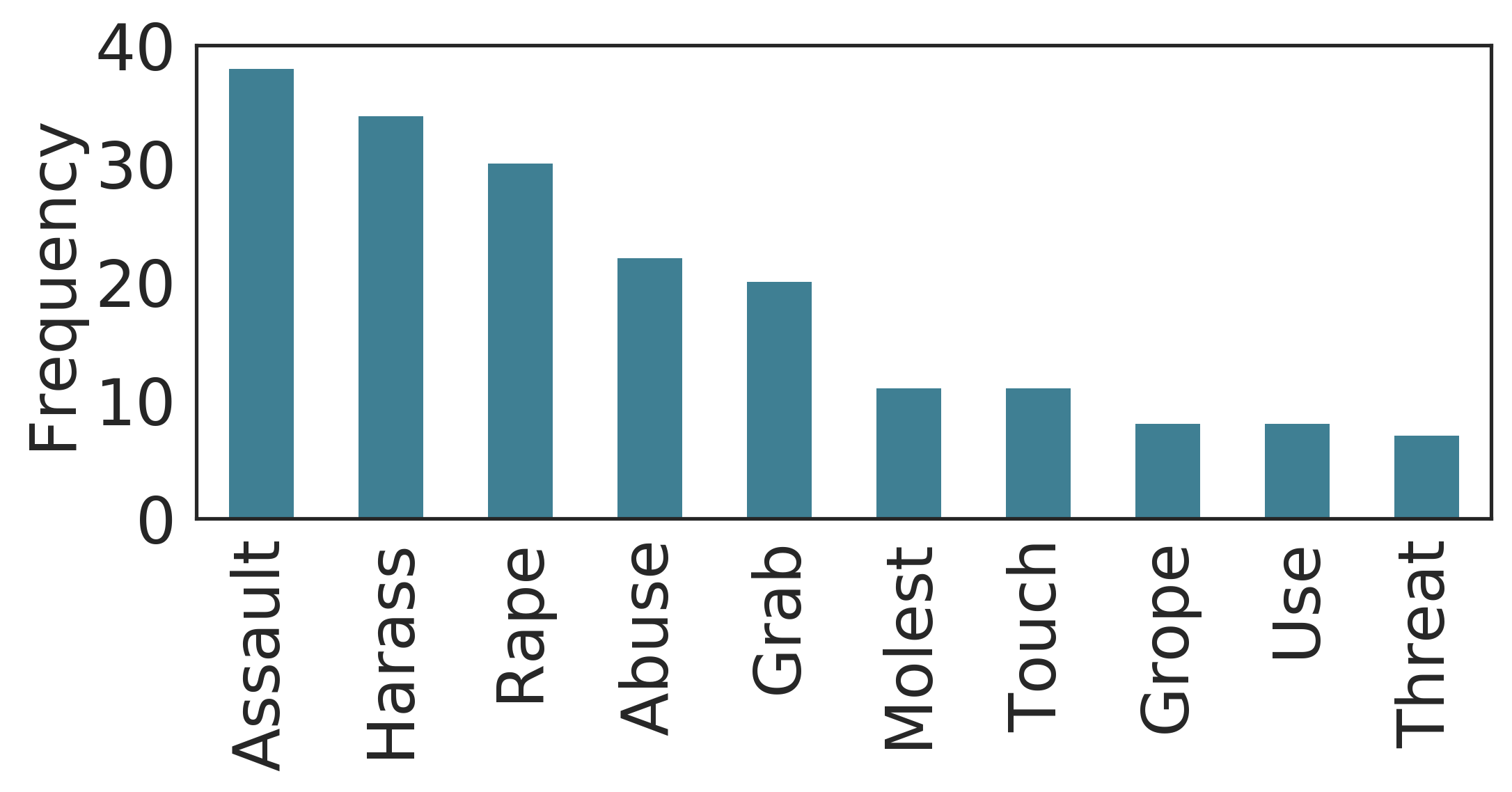}
  \caption{U.S. Cities}
  \label{fig-verb-frequency-us}
\end{subfigure}
\vspace{-0.12in}
\caption{Frequency of top-$10$ harassment categories in Asian cities (a) and in U.S. cities (b).}
\label{fig-verb-frequency-continent}
\vspace{-0.12in}
\end{figure}

\begin{figure}[h]
\centering
\includegraphics[width=0.9\linewidth]{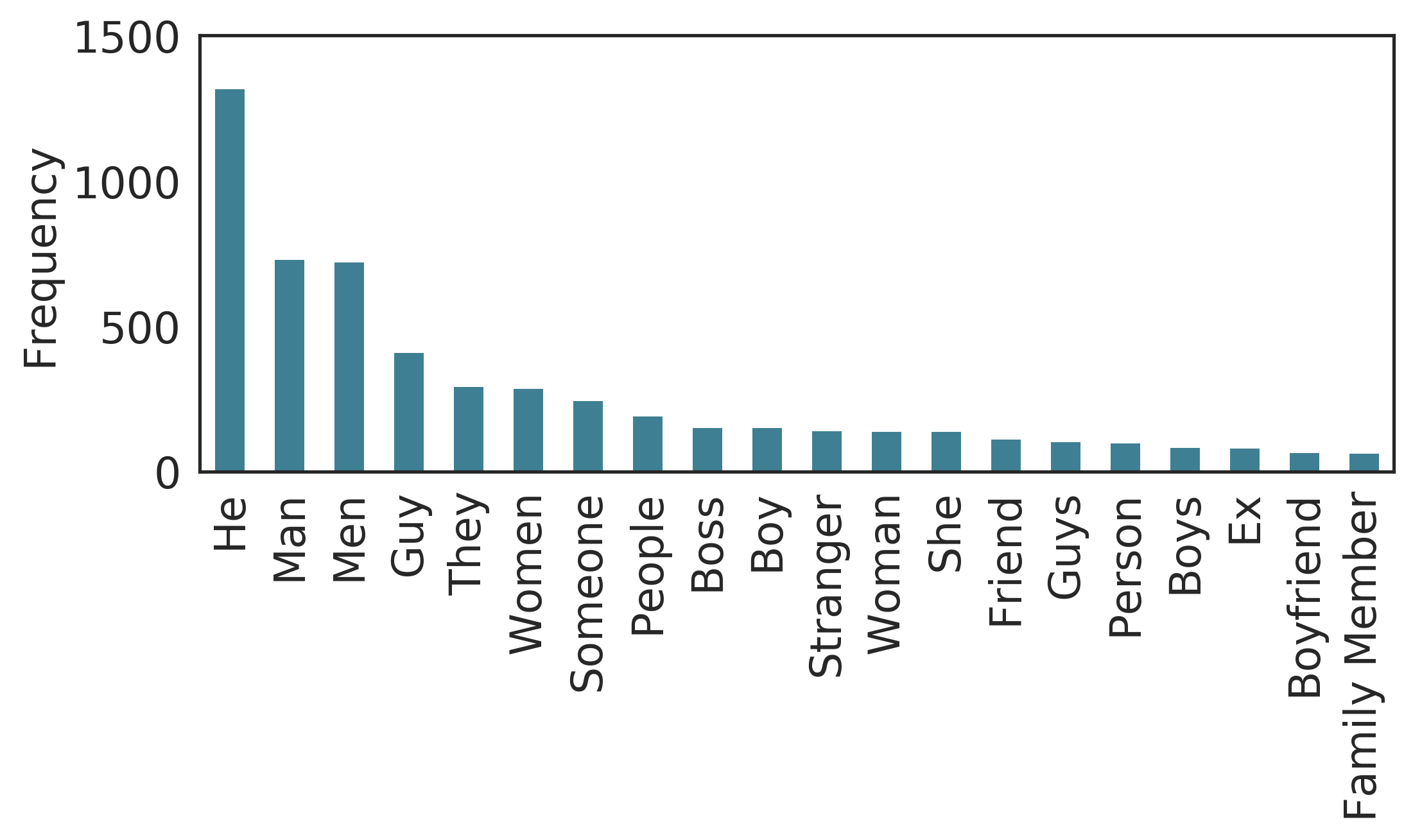}
\vspace{-0.3in}
\caption{Frequency of harasser roles.}
\label{fig-harasser-frequency}
\vspace{-0.12in}
\end{figure}

To detect the harassment related claims, we looked at the top-$1000$ most frequent verbs in the dataset and identified a set of $27$ verbs, $\mathcal{H}$, which are generally associated with sexual harassment and similar events. The verbs are-- \emph{Abuse, Assault, Attack, Beat, Bully, Catcall, Flirt, Fondle, Force, Fuck, Grab, Grope, Harass, Hit, Hurt, Kiss, Masturbate, Molest, Pull, Rape, Rub, Slap, Stalk, Threat, Touch, Use, Whistle}. From now on, we use the term \emph{verb} and \emph{category} interchangeably. We understand that the presence of a harassment related verb in a claim does not necessarily indicate a confession of a harassment event. So, to avoid false positives, we acted conservatively and only considered the claims where the \emph{verb} $\in \mathcal{H}$, the \emph{subject} is either a \emph{noun} or in third person (\emph{she, he, they, etc.}), and the \emph{predicate} exists (i.e., not an empty string). In total, we identified $15,585$ such claims. Table ~\ref{tab-example} shows some examples of the claims and their corresponding tweets. It should be noted that the SRL detection technique identifies presence of negation of an action (e.g., \emph{He did\textbf{n't} kiss her}) or modal of an action (e.g., \emph{Women \textbf{can} abuse men}). We omitted the claims containing negation and modal from our analysis. Figure ~\ref{fig-verb-frequency} shows frequency of top-$20$ most frequent harassment categories. About $60\%$ of all harassment claims belong to the top-$5$ categories. There are $6,445$ unique harasser roles among these claims. Figure ~\ref{fig-harasser-frequency} shows the top-$20$ most frequent roles along with their frequency in the harassment claims. One interesting observation is that in most cases, the victims did not disclose any information about the attacker and rather referred as \emph{He}. In many cases, the victims used terms such as \emph{men}, \emph{man}, or \emph{Boy} which give some indication about the age of the harasser. To understand the extent of harassment in different parts of the world, we plot the frequency of top-$10$ harassment categories in $4$ Asian cities (figure ~\ref{fig-verb-frequency-asia}) and in $7$ U.S. cities (figure ~\ref{fig-verb-frequency-us}). The pattern is almost the same in these two continents. $8$ categories are common among the top-$10$ categories of both continents. 

To further understand the \emph{harassment category -- harasser role} interplay, we draw a heatmap of these two categorical variables. Figure ~\ref{fig-verb-frequency} shows frequency of harassment claims pertaining to top-$20$ categories and top-$40$ harasser roles. We should note that not all categories are equal. The higher frequency of the words ``harass" and ``assault" than the actual description of the incident, such as ``grab", ``rape", ``kiss", ``rub", indicates that many women may still be uncomfortable to remember and share their harassment incident. They were hiding the actual even under the blanket words like `assault', 'harassment', or 'abuse'. It is well documented in literature that women often suffer from post traumatic stress disorders (PTSD) after being sexually harassed~\cite{avina2002sexual}, and thinking about a past even of harassment is also very painful for them~\cite{schauben1995vicarious}. Our finding reconfirms that fact. On the other hand, according to a study on sexual harassment typology ~\cite{gruber1992typology}, some of the categories such as \emph{Rape}, \emph{Fuck}, indicate higher level of severity and some are less severe such as \emph{Grab}, \emph{Kiss}, \emph{touch}. In our data, ``grab" is at number 3 just followed by ``rape", and ``fuck" is at number 20, which rejects any correlations between the severity of a harassment category and the number of tweets about it. This means, victims of all kinds of sexual harassment reported on twitter. Another interesting observation is that relatively minor sexual offenses like whistling, catcalling, offensive remarks, or disseminating behavior - which most women face in their everyday life~\cite{swim2001everyday} has not been reported much. One reason for this could be the fact that either many women have accepted those, or they found those to be too minor to report compared to the more serious harassment experiences that they have.



\begin{figure*}[ht]
\centering
\includegraphics[width=0.95\textwidth]{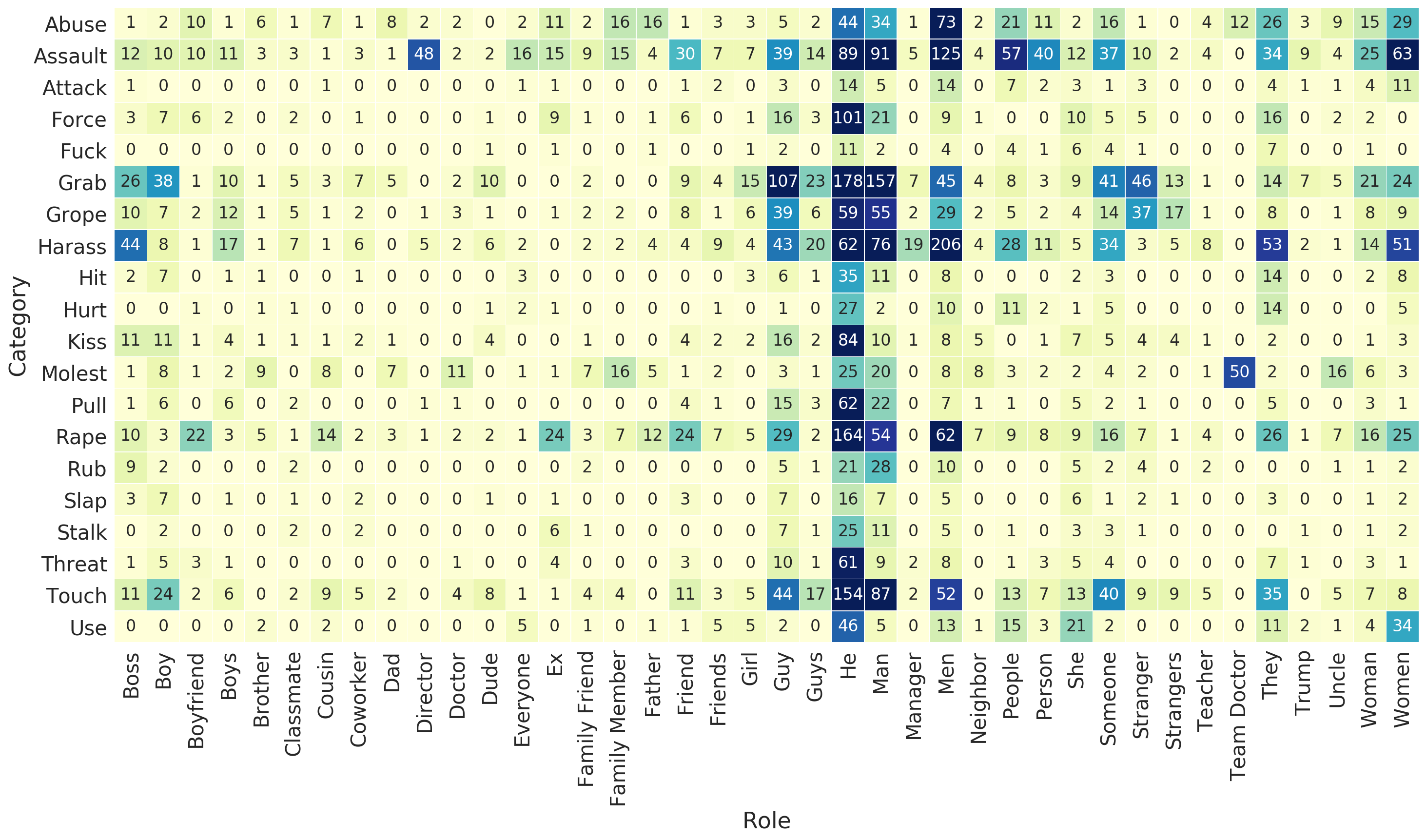}
\vspace{-0.2in}
\caption{Frequency distribution of \emph{harassment category -- harasser role} pairs.}
\label{fig-verb-harasser}
\end{figure*}

Next, we plot the frequency of each of the harassment categories against the nouns that occurred in the same tweets (see Figure ~\ref{fig-verb-harasser}). This graph reveals some interesting information. For example, characters like ``boss",``manager", ``teacher", ``coworker" were often alleged for ``harassing". If we look at the other categories of harassment by these characters, we find ``grab", ``grope", ``kiss", and ``touch", which reveal the nature of sexual harassment in workplaces. A person's relationships with these characters are often power-laden and work under an institutional setting. Hence, this observation indicates how these characters take the advantage of their institutional power to harass women, which supports the existing literature on workplace harassment ~\cite{russell1984sexual}. On the other hand, serious harassment incidents like ``rape" or ``molest" mostly came from a domestic or intimate relationship such as ``boyfriend", ``ex", ``friend", ``family member", ``uncle", and ``brother". These two observations together reveal a couple of facts. First, women are not safe either in their home or in their workplaces. Second, while people with a formal relationship are mostly engaged in milder harassment, men among a woman's familial or friend circles often commit severe harassment to her. These findings can be used to design laws, policies, and technologies to reduce sexual harassment in different context. 

Figure ~\ref{fig-verb-harasser} also shows how severe harassment incidents happen in specialized professional settings. For example, ``Team Doctor" was alleged for molesting by several posters, many of whom actually referred to the case of Dr. Larry Nassar. Dr. Nassar was a team doctor of USA Gymnastics and molested many young women by taking advantage of his profession ~\cite{larrynasser}. Similarly, ``Director" was accused of assaulting in many tweets, revealing a dirty face of film industry. ``Teacher" and ``Doctor" were accused of several harassment crimes. All these findings point to the lack of practice in professional ethics. Teachers and doctors, for example, help the sound development of a woman's body and mind. If a woman does not feel secure with them, then that will directly hamper their growth. Figure~\ref{fig-verb-harasser} also shows report of many ``strangers" or ``someone" harassing in various forms. This implies the ubiquity of this problem and the widespread vulnerability of women.

\begin{figure*}[th!]
\centering
\begin{subfigure}{0.17\linewidth}
  \centering
  \includegraphics[width=\linewidth]{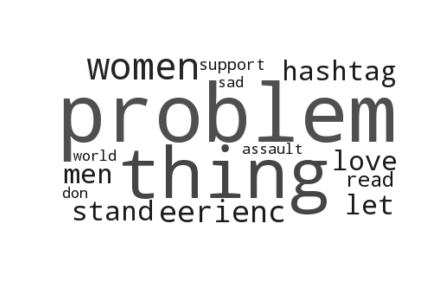}
  \vspace{-0.3in}
  \caption{Topic 1}
  \label{fig-topic-1}
\end{subfigure}%
\begin{subfigure}{0.17\linewidth}
  \centering
  \includegraphics[width=\linewidth]{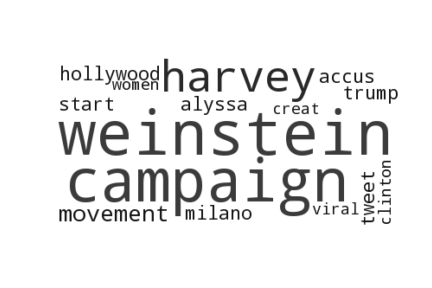}
  \vspace{-0.3in}
  \caption{Topic 2}
  \label{fig-topic-2}
\end{subfigure}%
\begin{subfigure}{0.17\linewidth}
  \centering
  \includegraphics[width=\linewidth]{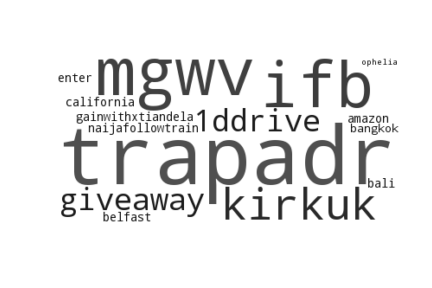}
  \vspace{-0.3in}
  \caption{Topic 3}
  \label{fig-topic-3}
\end{subfigure}%
\begin{subfigure}{0.17\linewidth}
  \centering
  \includegraphics[width=\linewidth]{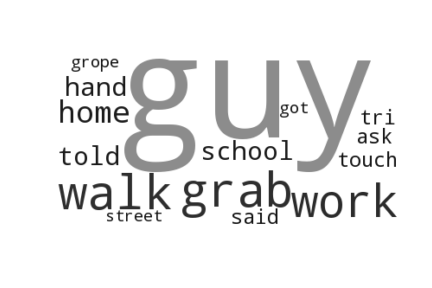}
  \vspace{-0.3in}
  \caption{Topic 4}
  \label{fig-topic-4}
\end{subfigure}%
\begin{subfigure}{0.17\linewidth}
  \centering
  \includegraphics[width=\linewidth]{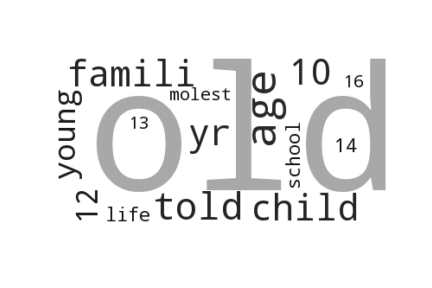}
  \vspace{-0.3in}
  \caption{Topic 5}
  \label{fig-topic-5}
\end{subfigure}%
\begin{subfigure}{0.17\linewidth}
  \centering
  \includegraphics[width=\linewidth]{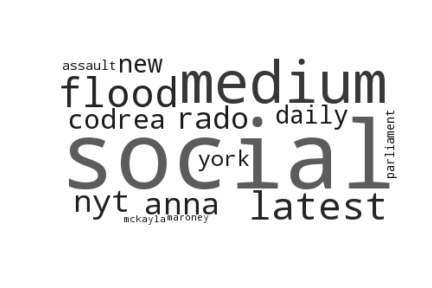}
  \vspace{-0.3in}
  \caption{Topic 6}
  \label{fig-topic-6}
\end{subfigure}
\vspace{-0.12in}
\caption{Topic distribution in the whole dataset.}
\label{fig-topics}
\end{figure*}


\subsection{Topic Distribution}
To understand the primary topics that surfaced during the \emph{\#MeToo} movement, we performed topic modeling over the whole corpus. One concern regarding traditional topic modeling algorithms (e.g., Latent Dirichlet Allocation, Latent Semantic Analysis) was, they emphasize on document-level word co-occurrence patterns to discover the topics of a document. So, they may struggle with the high word co-occurrence patterns sparsity which becomes a dominant factor in case of shorter context such as in tweets. 
To overcome this challenge, we used Biterm Topic Modeling (BTM) ~\cite{yan2013biterm} which generates the topics by directly modeling the aggregated word co-occurrence patterns of a short document. We empirically set the topic number as $6$. Each topic is represented by a set of words. Figure ~\ref{fig-topics} shows the distribution of topics over the whole dataset. We only show the top-$15$ most relevant words for each topic. Relevance of a word with respect to a topic, $r(w, k)$,  is calculated using an equation which is proposed in ~\cite{ldavis}-

$$r(w, k) = \lambda*\log(\sigma_{kw}) + (1-\lambda)*\log(\frac{\sigma_{kw}}{p_{w}})$$

Here, $\sigma_{kw}$ denotes the probability of word $w$ for topic $k$, $p_{w}$ denotes the marginal probability of term $w$ in the whole dataset, and $\lambda$ is a tuning parameter. We set $\lambda = 0.6$, as suggested in ~\cite{ldavis}. We used two visual channels, size and lightness, to encode the relevance of a word in Figure ~\ref{fig-topics}. Overall, the topics capture different facets of the \emph{\#MeToo} movement. Topic 1 captures the recognition of the sexual harassment as a problem and the supportive emotions expressed towards the overall movement. Topic 2 primarily indicates the \emph{Harvey Weinstein Scandal} that commenced the \emph{Weinstein Effect}~\footnote{{https://en.wikipedia.org/wiki/Weinstein\_effect}}. Topic 3 captures various advertisement efforts that attempted using the trending \emph{\#MeToo} hashtag to attract more traffic. Topic 4 and 5 exhibit some characteristics of the harassment claims. For instance, words like (\emph{grope}, \emph{grab}, \emph{touch}), and (\emph{street}, \emph{school}, \emph{work}) in topic 4 indicate the nature and location of the harassing events, respectively. In topic 5, words such as \emph{child}, \emph{young}, \emph{molest}, \emph{famili (stemmed form of family and related other words)}, and the numbers (e.g., $12$, $13$, $14$, $16$) insinuate many harassment claims where the victims were considerably very young. The last topic mostly comprises of the media coverage of the movement.


\subsection{Psychological Constructs Analysis}
We use widely known Linguistic Inquiry and Word Count (LIWC)~\cite{pennebaker2015development} dictionary to perform psychological constructs analysis of this dataset. Given a text, LIWC can perform a word level analysis on various psychological processes. In this experiment, we pick \texttt{Affective process}, \texttt{Biological process} and \texttt{Drives}. Within affect, LIWC can categorized a word into \textit{positive emotion, negative emotion, anxious, sad} and \textit{anger} bucket. Such analysis will allow us to understand the true sentiment (positive or negative) of tweets written by various users along with their emotional standpoint~ (angry, sad or anxious). In the biological process and drives category, we pick \textit{body, sexual} and \textit{power, risk, reward}, respectively as sub-categories to analyze. By categorizing a word into a biological process we can observe how vivid (biologically) an author of a tweet is when describing a \#MeToo experience. Similarly analyzing word distribution of drives category help us to establish a well known drive theory~\cite{berlyne1960conflict}, understanding what makes~(power, risk, reward) an author of a tweet to share a personal \#MeToo experience or showing positive/negative standpoint with the movement.

\begin{figure}[t!]
\centering
\begin{subfigure}{0.5\linewidth}
  \centering
  \includegraphics[width=\linewidth]{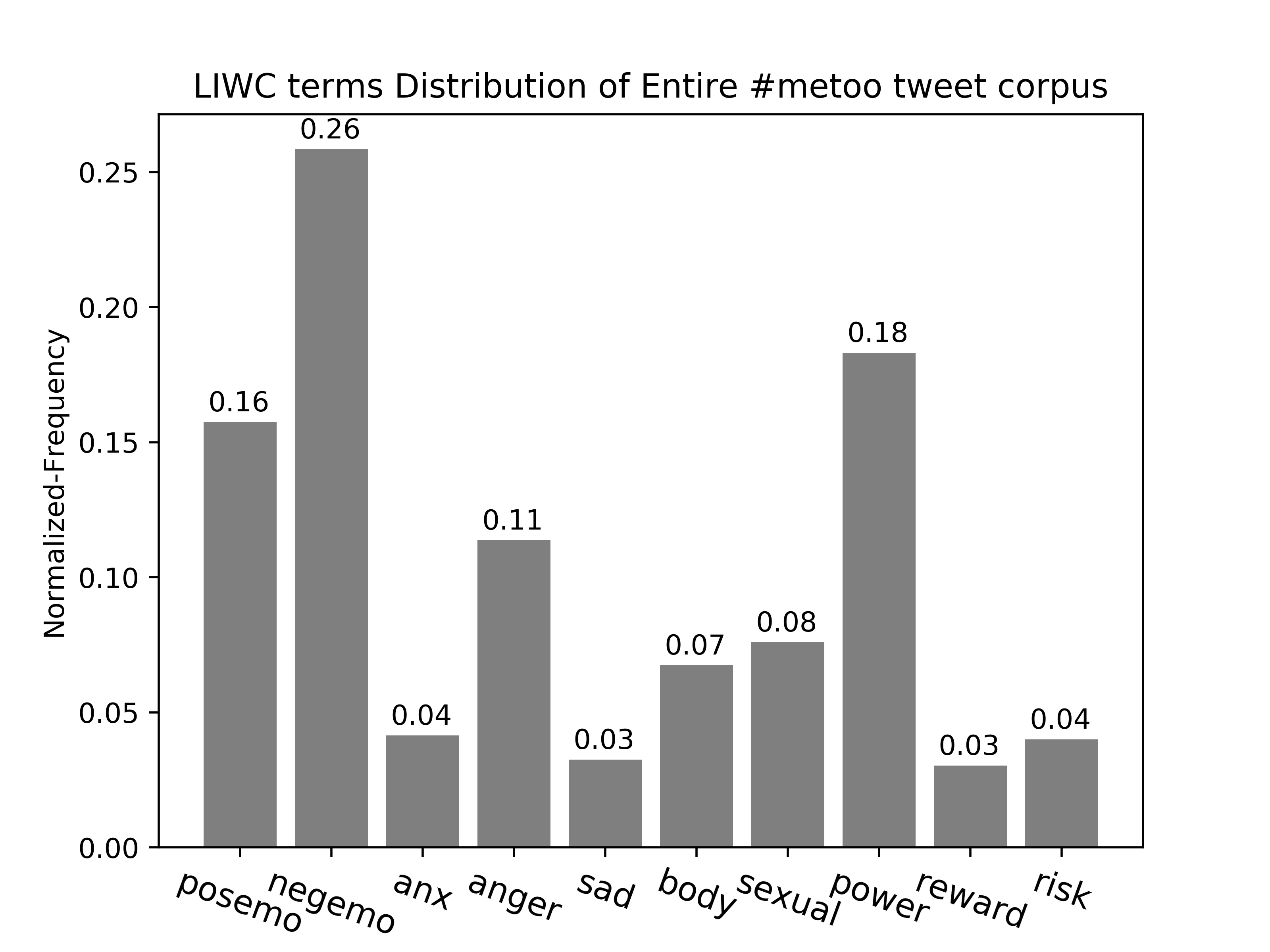}
  \vspace{-0.15in}
  \caption{Entire \#MeToo tweet Corpus}
  \label{fig-LIWC-all}
\end{subfigure}
\\
\begin{subfigure}{0.5\linewidth}
  \centering
  \includegraphics[width=\linewidth]{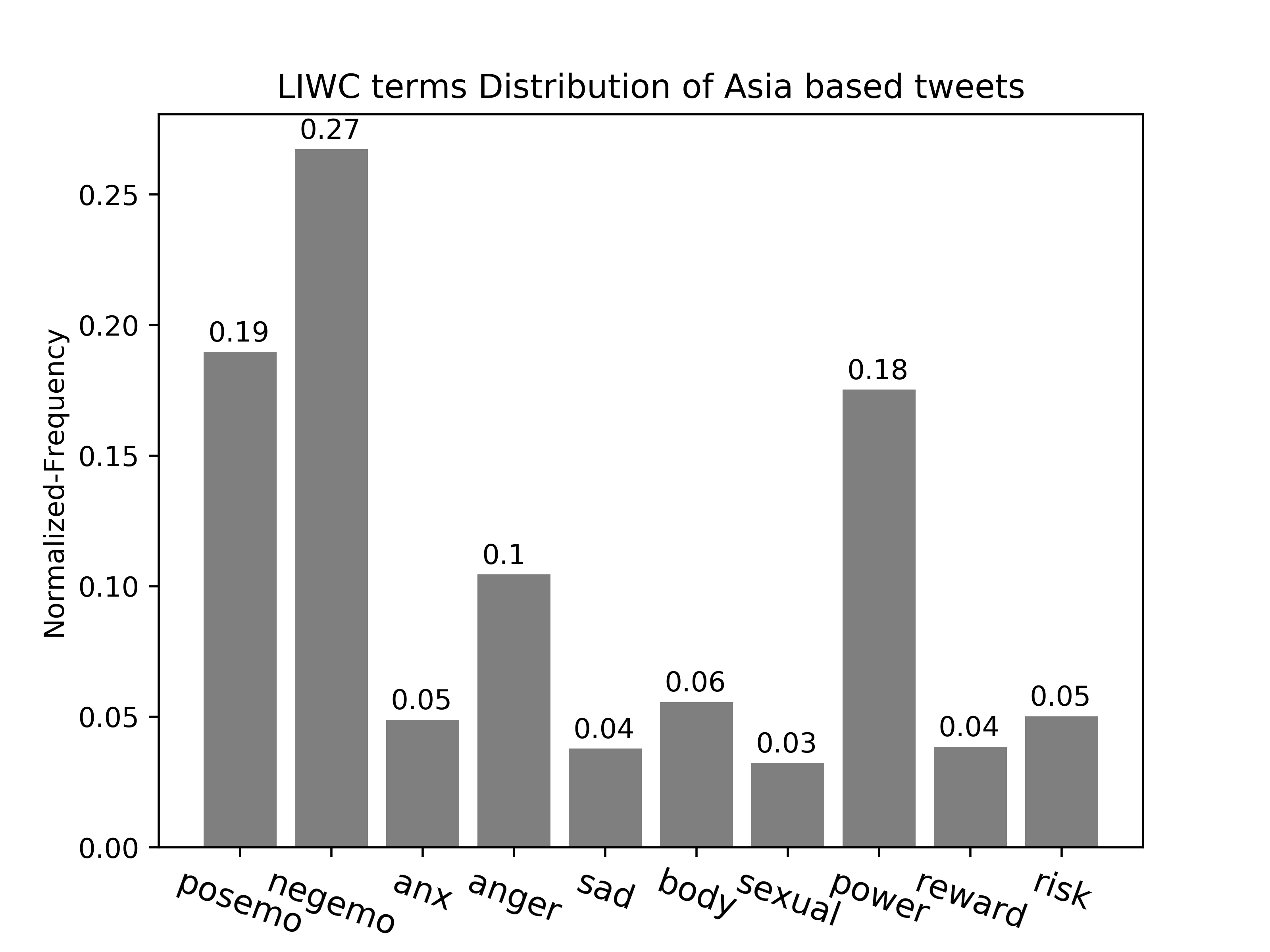}
   \vspace{-0.15in}
  \caption{Asian Cities}
  \label{fig-LIWC-asia}
\end{subfigure}%
\begin{subfigure}{0.5\linewidth}
  \centering
  \includegraphics[width=\linewidth]{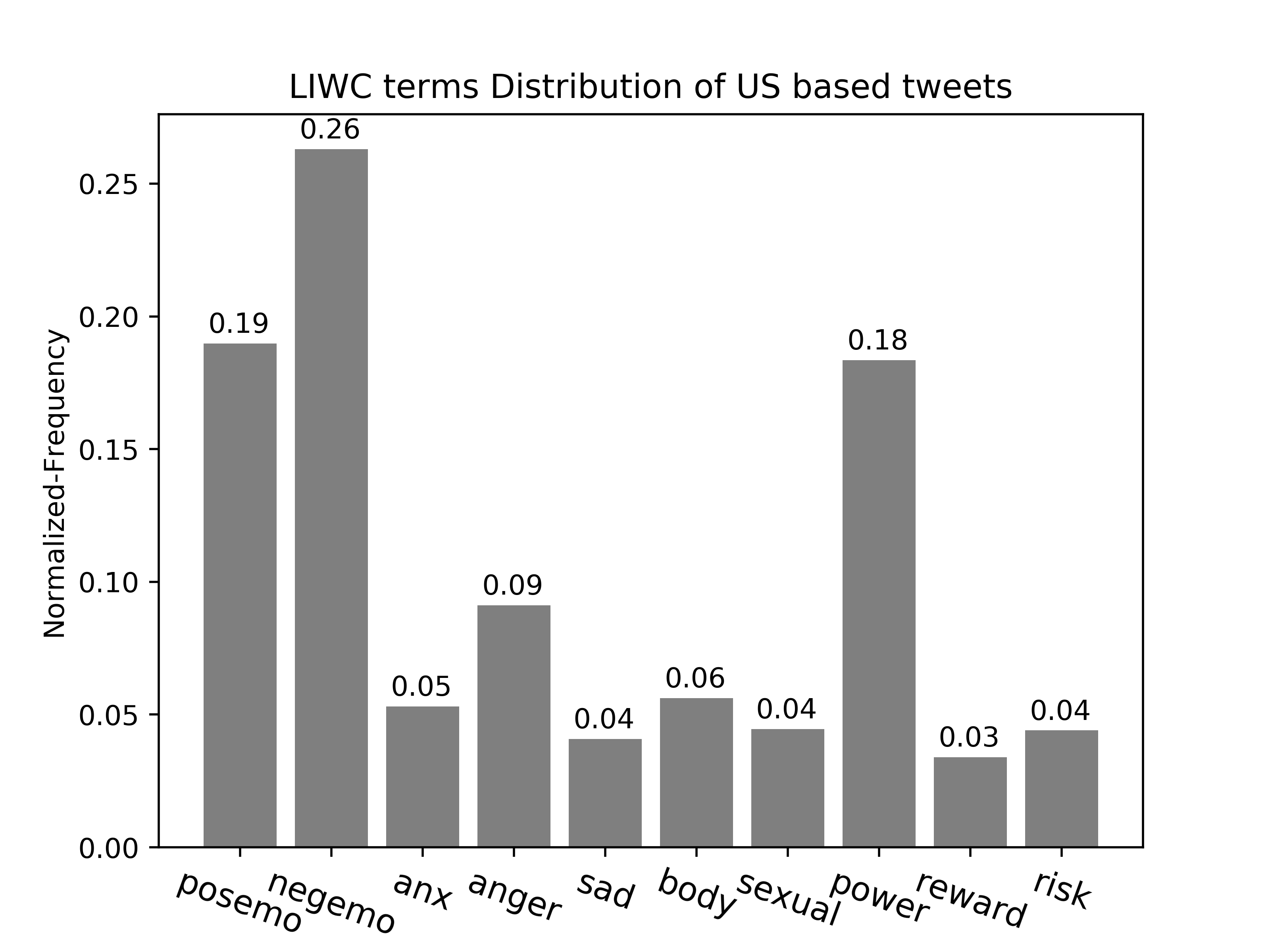}
   \vspace{-0.15in}
  \caption{U.S. Cities}
  \label{fig-LIWC-US}
\end{subfigure}
\\
\begin{subfigure}{0.5\linewidth}
  \centering
  \includegraphics[width=\linewidth]{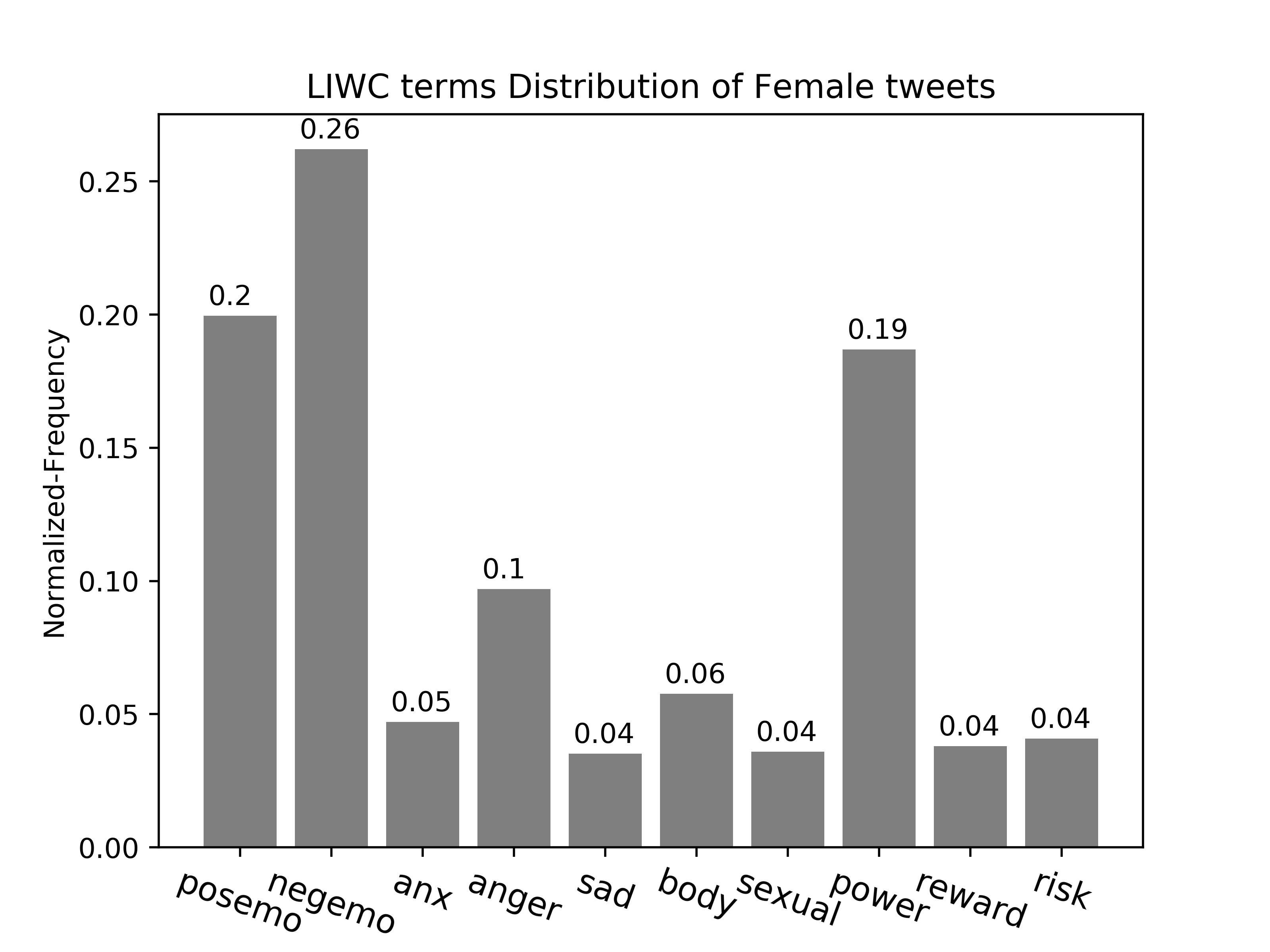}
   \vspace{-0.15in}
  \caption{Tweets from Females}
  \label{fig-LIWC-female}
\end{subfigure}%
\begin{subfigure}{0.5\linewidth}
  \centering
  \includegraphics[width=\linewidth]{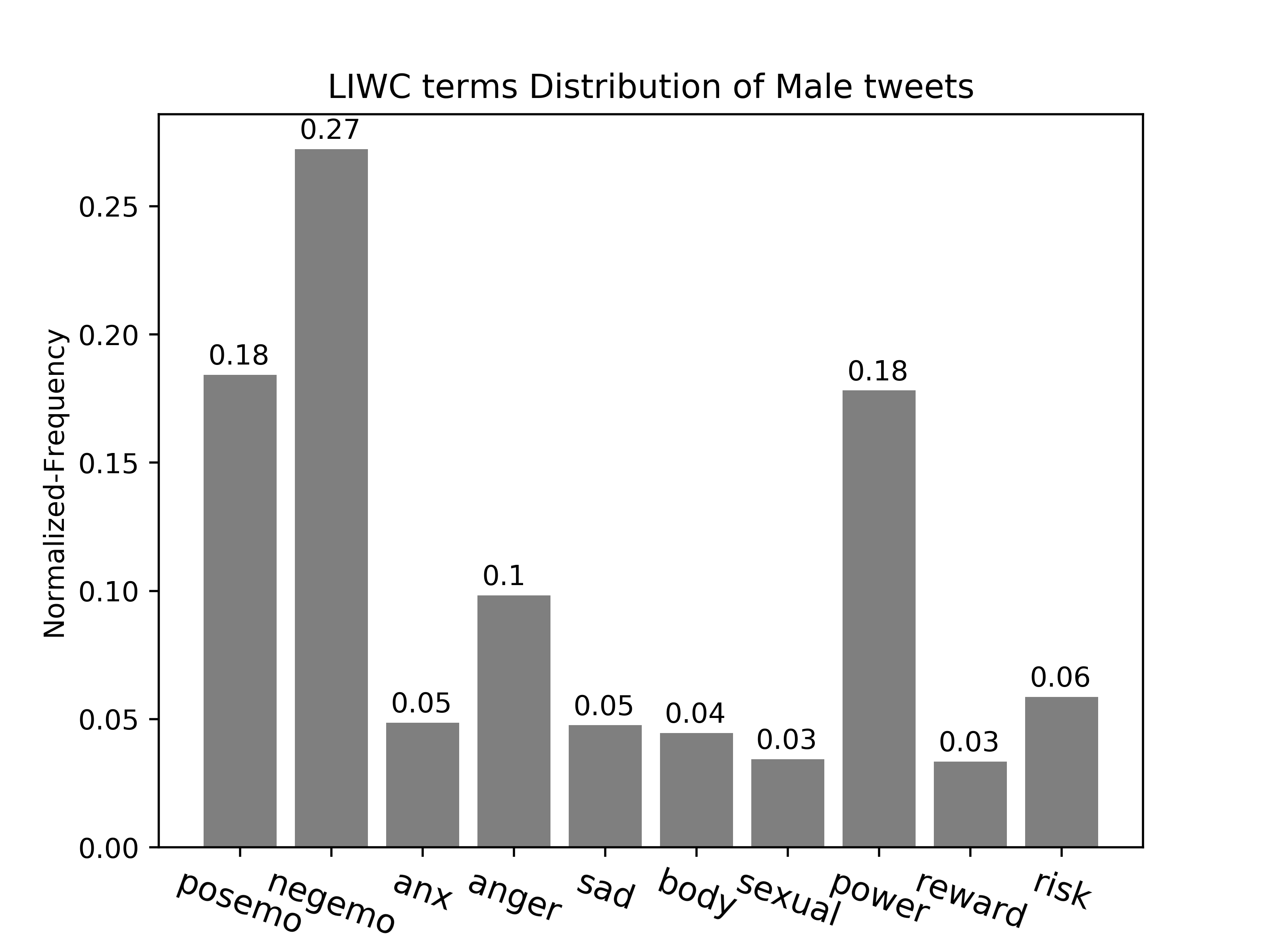}
   \vspace{-0.15in}
  \caption{Tweets from Males}
  \label{fig-LIWC-male}
\end{subfigure}
 \vspace{-0.15in}
\caption{LIWC term distribution}
\label{fig-LIWC}
 \vspace{-0.15in}
\end{figure}

We perform 5 different experiments. First we compute the word distribution of the sub-categories mentioned above in the entire dataset. We found the overall coverage of LIWC dictionary in this dataset to be approximately 50\%. Next, we extract tweets based on continent~(U.S. and Asia) and Gender (Male, Female) and compute similar word distribution using LIWC. In Figure~\ref{fig-LIWC}(a-e), we show the normalized frequency LIWC terms for each category. 

Figure~\ref{fig-LIWC-all} shows that the frequency of negative emotion is much higher than positive which establishes the overall sentiment of the corpus. In emotional standpoint majority of the tweets are expressing anger than sadness or anxiousness. Between sexual and body, sexual expression is slightly higher than the body. Within drives, power is significantly higher than reward or risk. This indicates exercising power was the major driving factor in this movement than some reward or risk.

In Figure~\ref{fig-LIWC-asia} and ~\ref{fig-LIWC-US}, we compute word distributions of tweets from Asia and US continent. As we can see, negative emotion is the majority in both but slightly higher is Asia-based tweets. Anger is expressed more than sadness and anxiousness in both cases. Biological process categories have similar patterns. Exercising power is the main driving factor in both cases but risk factor and reward is slightly higher in Asia-based tweets than US.

In Figure~\ref{fig-LIWC-male} and ~\ref{fig-LIWC-female}, we compute the same distribution as above of tweets from male and female authors. Interestingly, negative emotion is expressed slightly higher in male authors' tweets whereas positive emotion is higher in female authors' tweets. Anger is prominent in both the cases. Biological processed are higher in female tweets and this observation is intuitive as \#MeToo movement is centered on female. Once again, exercising power is the major driving factor in both cases.

These graphs serve as good indicators of a number of emotional constructs in the society. For example, most feminist scholars have identified sexual harassment as a result of power imbalance in the society ~\cite{hester1996women}. Most of them have opined that sexual harassment often has the primary objective of establishing the `ownership' and `dominance' in a symbolic way ~\cite{uggen2004sexual}. Our analysis over the large amount of tweets reconfirms those opinions. Second, the emotional responses from the male posters indicate the role of social media platform to transfuse emotions from one community to another. Although most of these male posters were never sexually harassed, they could connect with the female posters emotionally. We argue that such emotional connections are important to create a wide awareness against sexual harassment. Third, the negligible difference in the nature of responses to sexual harassment between Asia and US continent demonstrates the hidden sadness of women across the globe. We often define development by economic stability, infrastructural development, and technology innovation, and thus some parts of the world becomes more ``developed" than the others. However, when we take a closer look at the internal emotion of people, especially the women, we find similar kind of sadness that bind them together, and challenge the definition of development.





%% file: sec-limitations.tex
\section{Limitations}

Although the findings of this paper reveal a number of very interesting and important facts around gender, harassment, and movements on social media, all these findings are based on our dataset and hence they may not be generalized. Our dataset consists of only the \#MeToo hash-tagged tweets that were posted between October 15-31, 2017. Hence, this dataset does not contain any tweets that were posted later, and hence, our dataset does not represent the whole \#MeToo movement on Twitter. However, it should be noted that the movement was at its peak during that period of time. Hence, although our data does not represent the entirety of \#MeToo movement, it does represent a significant portion of it. Second, \#MeToo movement also took place in other social media including Facebook and Instagram, and we have not analyzed those data. Although we believe that the findings from our data will not be challenged by the posts on those platforms, we may miss some important themes because Twitter allows short text only. Third, all the findings that we have found in our data are actually based on how people talked about sexual harassment with \#MeToo on twitter, and these may not represent the actual reality. Several studies show that people often lie, cheat, and misrepresent themselves on social media for various purposes (see ~\cite{mallan2009look}, for example). However, to minimize this effect, we have tried our best to triangulate our findings with existing scholarship around related topics. Overall, although this dataset is not complete and it suffers from many shortcomings that most analyses of online data have, the findings of this paper represent some dominant patterns of sexual harassment and online movements that are prevalent across the globe. 

\label{sec-limitations}

%% file: sec-discussion.tex
\section{Concluding Discussion}
In this paper, we have presented our quantitative analysis of a large set of tweets with \#MeToo hashtag. We have shown different interesting patterns in the data and discussed their possible implications. Our analysis thus contributes to the studies of sexual harassment, social media, online movement, and feminism in general. Along with these contributions, this analysis generates answers to two important research questions that this paper asked - 

\textbf{Who and What:} First, our data shows that women from all over the world participated in \#MeToo online movement, and their harassment reports are more similar than different. Most of them reported about severe harassment incidents and skipped the minor ones. Their tweet contained their sadness and the consequences of power exercise. Their emotions also transfused to many men over Twitter, and they also posted with \#MeToo and expressed their solidarity with women. At the same time, we have seen many other progressive movements joined \#MeToo movements because of their political benefits. Finally, many people or organizations used these hash-tag to get attention and promote their products or services.

\textbf{How and by Whom}: Second, our data also demonstrates that while sexual harassment is pervasive, there is still some dominant patterns. For example, women are often severely harassed (raped, molested) by the people in their close family or friend circles or by their intimate partners. At the same time, they are being victims of minor harassment (grab, grope, touch) at their workplaces. Similarly, people in different specialized professions (doctors, teachers) are often taking advantage of their job to harass women severely. Besides these, women are often being harassed by complete strangers. These patterns point to the vulnerable position of women in the society, where they are often subject to power exercises. 

\label{sec-discussion}